\documentclass{ws-p8-50x6-00}

\begin{document}

\title{Three-nucleon portrait with pion}

\author{L. Canton and G. Pisent}

\address{INFN and Dipartimento di Fisica dell'Universit\`a di Padova
         \\
         Via F.Marzolo 8, Padova I-35131, Italy} 

\author{W. Schadow}

\address{TRIUMF, 4004 Wesbrook Mall, Vancouver, BC, Canada V6T 2A3}

\author{T. Melde and J. P. Svenne}

\address{Physics Department, University of Manitoba and 
Winnipeg Institute for Theoretical Physics \\
 Winnipeg, MB, Canada R3T 2N2}


\maketitle

\abstracts{We report on recent results obtained by the above collaboration
on the collision processes involving three nucleons, where we pay
particular attention on the dynamical role of the pion.
After discussing the case at intermediate energies, where real pions
can be produced and detected, we have considered the case at lower energies,
where the pions being exchanged are virtual. The study has revealed 
the presence of some new pion-exchange mechanisms, which leads
to a new three-nucleon force of tensor structure.
Recently, the effect of this tensor three-nucleon force 
to the spin observables for neutron-deuteron scattering
at low energy has been analyzed, and will be briefly reviewed.}

\section{Introduction}

A crucial aspect in modern three-nucleon dynamics concerns the inclusion
in the system of additional mesonic aspects which cannot be conveniently
described by a conventional ({\it e.g.}, meson-exchange) two-nucleon potential.
The standard approach to this problem introduces three-nucleon forces (3$NF$),
which effectively describe the addition of such irreducible mesonic 
contributions.
Such ``3$NF$'' approaches turned out to be successful in curing
the problem of the underbinding of the three-nucleon 
bound state, but failed to explain the puzzle of the vector
analyzing powers (``the $A_y$ puzzle'') in nucleon-deuteron scattering
at low energy.

Recently, with the aim to explain the $A_y$ puzzle,
there have been various approaches which introduced 3$NF$ 
of new structure, in the attempt to characterize, 
more or less effectively,  some refined aspects of the 
meson dynamics which were not yet contemplated in the 
traditional 3$NF$ expressions.
The starting point of these descriptions is characterized 
by a Hamiltonian which is restricted in the three-nucleon space,
wherein the mesonic aspects have been integrated out from the very beginning
via the introduction of 3$NF$-like contributions.

In contrast, we have considered an approach 
which starts from the employment of the full 
four-body dynamics of the one-pion three-nucleon
system. The  method is based on a generalization 
to the one-pion three-nucleon problem of the rigorous 
four-body theory of Grassberger-Sandhas-Yakubovsky\cite{GSY}, 
this generalization being developed by one of the
authors\cite{Canton98}. Such generalization 
turned out to be a highly nontrivial task,
and a physically sound approximation scheme has been recently 
derived for the resulting dynamical equations\cite{CMS}.

To the lowest order, such approximation scheme reproduces the diagrams
leading to the 3$NF$ {\it \'a la} Tucson-Melbourne\cite{TM}. However, at
the same level of approximation, other irreducible diagrams of
different structure appeared.  In particular, a recent detailed
analysis\cite{CSa} for one class of such diagrams showed the
appearence of a new three-nucleon force of tensor structure.  This 3$NF$
component is generated by the underlying one-pion exchange diagram
where one of the two nucleons interacts with the third one while the
pion is being exchanged.  In a conventional 3$NF$ approach, where the
meson degrees of freedom are integrated out from the very beginning,
such diagram is usually neglected because of the presence of a
cancellation effect involving meson-retardation effects of the
combined exchange of two pions.  However, by treating explicitly the
pion dynamics, it was found\cite{CSa} that this cancellation is
incomplete, and a 30\% effect survives once these mesonic retardations
are taken into account.  In a recent work\cite{CSb}, the effect of
this new 3$NF$ of tensor structure has been studied in neutron-deuteron
scattering at low energy.

\section{Pion production from few-nucleon collisions}

Here we summarize the results obtained on pion
production from collisions involving few-nucleon systems.
For full details, we refer to the original 
papers$^{7-11}$.
The starting point is the low-energy Lagrangian
coupling the pion ($\Phi $) and nucleon ($\Psi $) fields,
\vspace{4pt}
\begin{eqnarray}
\label{Lagrangiana}
{\cal L}_{\rm int}
& = & 
{f_{\pi NN}\over m_\pi} 
\bar\Psi\gamma^\mu\gamma^5\vec\tau\Psi \cdot \partial_\mu \vec \Phi \\
&&
-4\pi{\lambda^{}_I\over m_\pi^2}
\bar\Psi\gamma^\mu\vec\tau\Psi \cdot 
\left[\vec\Phi \times \partial_\mu \vec \Phi\right] 
-4\pi{\lambda^{}_O\over m_\pi}
\bar\Psi \Psi 
\left[\vec\Phi \cdot \vec \Phi\right]\ . \nonumber
\end{eqnarray}
The choice of the two constants $\lambda^{}_I$ and $\lambda^{}_O$
is consistent with the pion-nucleon scattering lengths; however
the kinematics for the  pion-production mechanism 
requires an off-shell extrapolation of the two constants, which can
be conveniently obtained by representing the corresponding four-leg
vertices in terms of a combination of exchanges of heavier mesons, 
$\rho$, $\sigma$ and some additional effective short-range 
effects\cite{Hernandez96}.

\begin{center}
\begin{figure}[t]
%\figurebox{20pc}{10pc}{} % to have a box alone
\epsfxsize = 27pc 
% will enlarge or reduce the postscript figures based on the xsize
\epsfbox{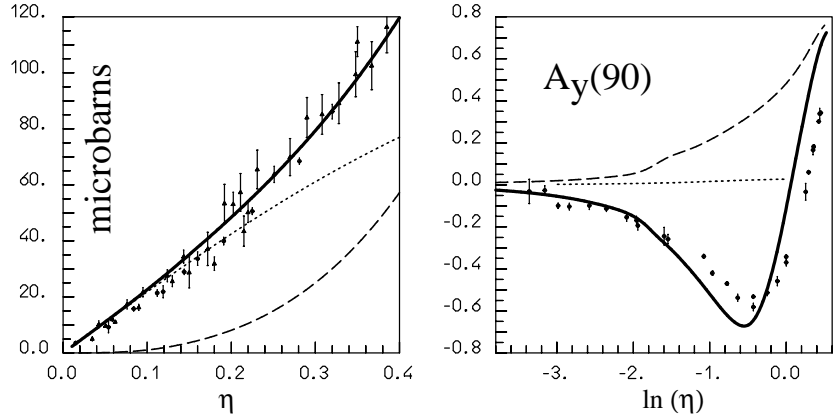} % postscript image file name
\caption{Total cross section ({\it left}) and proton analyzing power
({\it right}) at $90^o$ for the $pp\rightarrow \pi d$ reaction. 
$\eta$ is the dimensionless pion c.m. momentum in units of pion masses.
The solid line corresponds to the complete model. The dotted line
corresponds to the mechanisms implied by the low-energy Lagrangian
of Eq.~(\ref{Lagrangiana}). The dashed line represents the $p$-wave
($\pi$N) mechanisms and includes the $\Delta$ effects. 
The data shown are Coulomb corrected experimental data, which are 
referenced elsewhere \protect \cite{cdd98}. }
  \label{fig:1}
\end{figure}
\end{center}

The description of the pion production process in terms of this 
Lagrangian is however not sufficient, because of the strong coupling 
of the $\pi$N system to the $\Delta$ channel. 
Therefore, one must consider, in addition,  the mechanisms triggered
by the $\pi$N$\Delta$ coupling and by the $V(\Delta N-NN)$ transition 
potential. As shown in Fig.~\ref{fig:1}, this occurs even
around threshold, since the structure of $A_y$ 
is governed by the interference effects between the
$\Delta$ and non-$\Delta$ mechanisms. The results 
obtained\cite{cs00,cpss00} for pion production from $pd$ 
collision show that this interplay is even more important here, 
and the relevance of the various production mechanisms shows up already
at the level of the integral cross section (see the contribution to these 
Proceedings by 
M.~Viviani,\cite{corto:mic.viv} Fig. 2). In particular,
the off-shell effects in the isoscalar channel
turned out to be very important, for the normalization
of the production cross-section, as well as for describing 
the spin observables at threshold.

\section{Pion dynamics and new three-nucleon-force diagrams}

%\begin{center}
\begin{figure}[b]
%\figurebox{20pc}{10pc}{} % to have a box alone
\epsfxsize = 15pc
\epsfysize = 4pc  
% will enlarge or reduce the postscript figures based on the xsize
\epsfbox{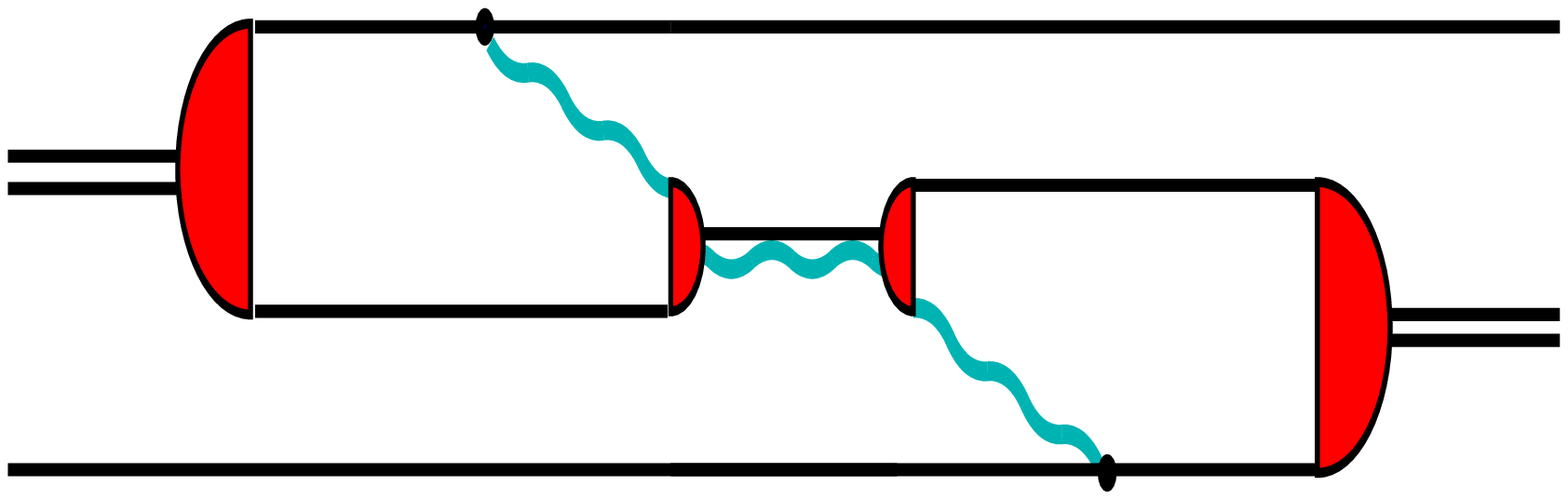} % postscript image file name
\epsfxsize = 12pc 
\epsfbox{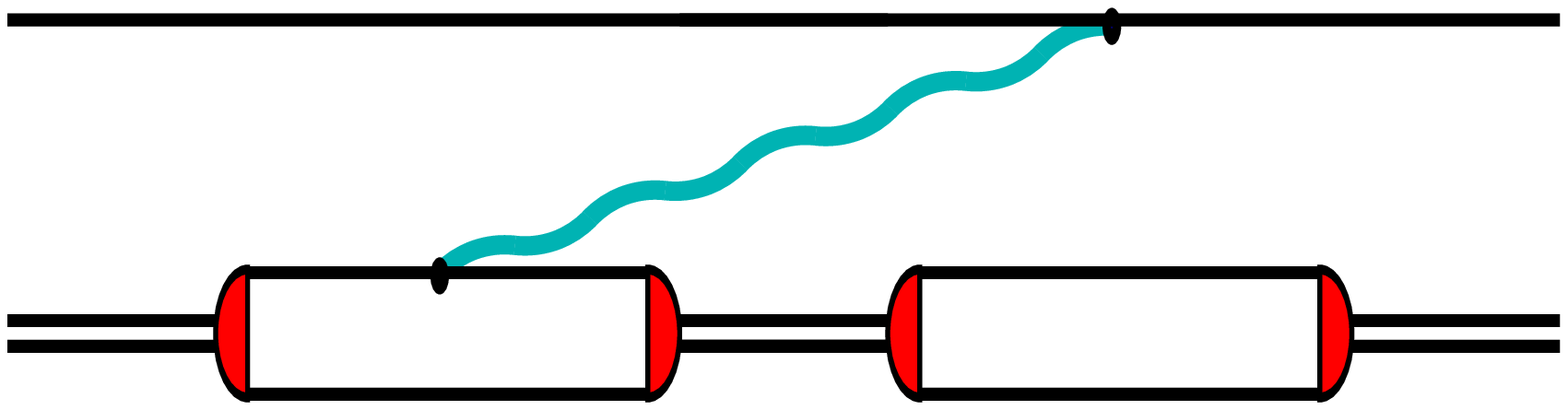} % postscript image file name
\epsfxsize = 20pc 
\centerline{\epsfbox{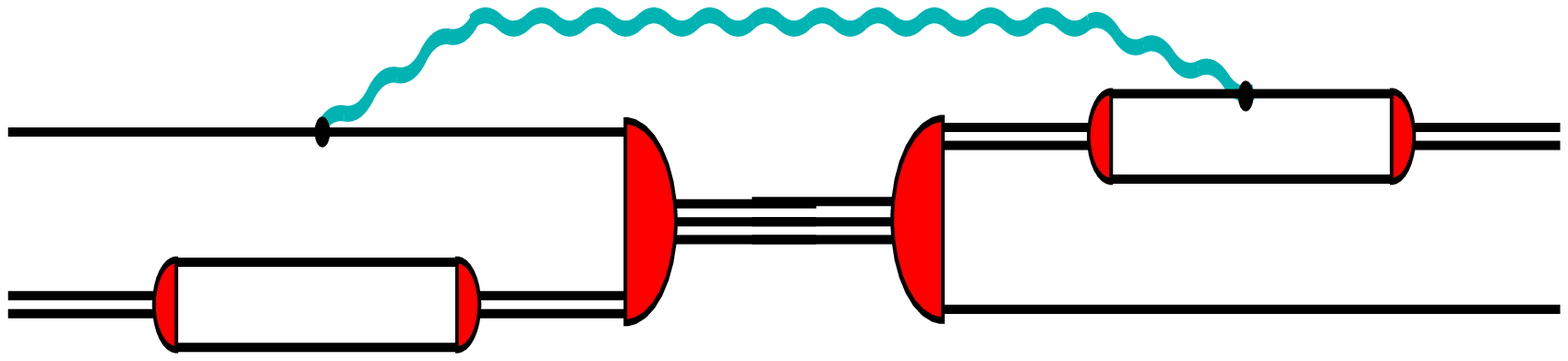}} % postscript image file name
\caption{Irreducible 3$NF$ contributions generated --at the lowest order-- by the pion dynamics in 
the AGS equation. The straight lines are nucleons,
the wavy line represents the pion.}
  \label{fig:2}
\end{figure}
%\end{center}

The explicit pion dynamics can be included in the 3$N$ system\cite{Canton98} 
by generalizing the rigorous 3$N$ approach in terms of AGS 
equations\cite{AGS}. One advantage is consistency
between the treatment of such mesonic aspects, and the
method of solution of the three-nucleon problem with a
given 2$N$ potential. Starting from this result 
a practical  approximation scheme has been derived\cite{CMS}, 
which treats perturbatively the four-body dynamics of the 
pion-three-nucleon system.
To the lowest order, the approach leads to 3$NF$ diagrams,
which can be incorporated into an effective two-cluster
3$N$ equation:
\begin{equation}
X_{ss'}^{(2)}=Z_{ss'}^{(2)}+\sum\limits_{s''}
{Z_{ss''}^{(2)}\tau_{s''}^{(2)}X_{s''s'}^{(2)}} \,  
\label{eLe}
\end{equation}
where the driving term is given by two contributions
 \begin{equation}
  Z_{ss'}^{(2)}=Z_{ss'}^{AGS}+Z_{ss'}^{3NF} \, .
  \label{Z-AGS-extended}
  \end{equation}
The first term represents the standard nucleon-exchange diagram
between different correlated pairs of nucleons, while the second
contains irreducible exchange diagrams involving 
the pion (see Fig.~\ref{fig:2}). Beside the standard
3$NF$ diagram --the first diagram in the figure--
where the pion rescatters with the nucleon
while being ``in flight'' 
(it represents the basic diagram for the construction of 
the ``standard'' three-nucleon force), 
we obtain at the same order two additional
diagrams which lead to 3$NF$ of different structure.
At the present stage we have considered the consequences
implied by the second diagram of Fig.~\ref{fig:2},
representing a 2$N$ rescattering process while the pion 
is ``in flight''. 
The combined effect of the two 3$NF$ terms implied by the first 
two diagrams in Fig.~\ref{fig:2}
have been studied also in a schematic one-dimensional 
model\cite{melde00}.
Further work is needed in order to consider
the last type of diagrams  shown in  figure.

\section{Tensor three-nucleon force}

The tensor structure of the 3$NF$ implied by the second diagram shown in 
Fig.\ref{fig:2} has been discussed recently\cite{CSa}. 
The explicit expression  is
\begin{eqnarray}
\label{OPE-$3NF$}
\lefteqn{V^{3N}_3({\bf p,q,p',q'};E) =
 {f_{\pi NN}^2(Q)\over m_\pi^2}{1\over (2\pi)^3}}
 \\ && \times
  \!\left [
{
({\mbox{\boldmath $\sigma_1$}}\cdot{\bf Q})
({\mbox{\boldmath $\sigma_3$}}\cdot{\bf Q})
({\mbox{\boldmath $\tau_1$}}\cdot {\mbox{\boldmath $\tau_3$}})
+
({\mbox{\boldmath $\sigma_2$}}\cdot{\bf Q})
({\mbox{\boldmath $\sigma_3$}}\cdot{\bf Q})
({\mbox{\boldmath $\tau_2$}}\cdot {\mbox{\boldmath $\tau_3$}})
\over
\omega_\pi^2
}
\right]  \nonumber \\
&&\times \, {
\tilde t_{12}({\bf p},{\bf p'};E-{{q}^2\over 2\nu} - m_\pi)
\over
2m_\pi}
\nonumber \\
&& +{f_{\pi NN}^2(Q)\over m_\pi^2}{1\over (2\pi)^3}
\, {
\tilde t_{12}({\bf p},{\bf p}';E-{{q'}^2\over 2\nu} - m_\pi)
\over
2m_\pi}
\nonumber   \\
& & \!\!
 \times \! \left [
{
({\mbox{\boldmath $\sigma_1$}}\cdot{\bf Q})
({\mbox{\boldmath $\sigma_3$}}\cdot{\bf Q})
({\mbox{\boldmath $\tau_1$}}\cdot {\mbox{\boldmath $\tau_3$}})
+
({\mbox{\boldmath $\sigma_2$}}\cdot{\bf Q})
({\mbox{\boldmath $\sigma_3$}}\cdot{\bf Q})
({\mbox{\boldmath $\tau_2$}}\cdot {\mbox{\boldmath $\tau_3$}})
\over
\omega_\pi^2
}
\right ] \, . \nonumber
\end{eqnarray}
The momenta ${\bf p,q}$ represent the Jacobi coordinates
of the pair ``$1$$2$'', and spectator ``$3$'' in the incoming state, 
and similarly  ${\bf p',q'}$ are for the outgoing channel. 
$E$ is the 3$N$ energy and ${\bf Q}$ is the momentum carried by the pion,
${\bf Q}={\bf q-q'}$.
The tensor structure arises because of the presence of the One-Pion-Exchange 
term between the spectator nucleon and the pair, while the pair
correlation is described by means of the 2$N$ $t$-matrix, evaluated
at the off-shell energy, as required from the kinematics for the 
subsystems.

The above expression has been obtained by taking into account
the dynamical effects of one single pion. It is clear, however,
that one should expect that the combined exchanges of more than
one pion will have some additional effects. 
However, the irreducible diagrams 
generated by the combined exchange of two pions --at leading order--
cancel out against the mesonic retardation effects of the 
reducible contributions --at second order-- of the Born diagram.
This cancellation has been observed by various authors\footnote{See, 
{\it e.g.}, the references contained here\cite{CSa}}, but
it has been also argued recently\cite{CSa} that this cancellation
is incomplete, and leads in Eq.~(\ref{OPE-$3NF$}) to a {\it subtracted}
$t$-matrix: 
$\tilde t_{12}({\bf p},{\bf p'};E-\dots)
=t_{12}({\bf p},{\bf p'};E-\dots)
-v_{12}({\bf p},{\bf p'})$. 
%\begin{center}
\begin{figure}[t]
%\figurebox{20pc}{10pc}{} % to have a box alone
\epsfxsize = 17pc \epsfysize = 9pc 
\centerline{\epsfbox{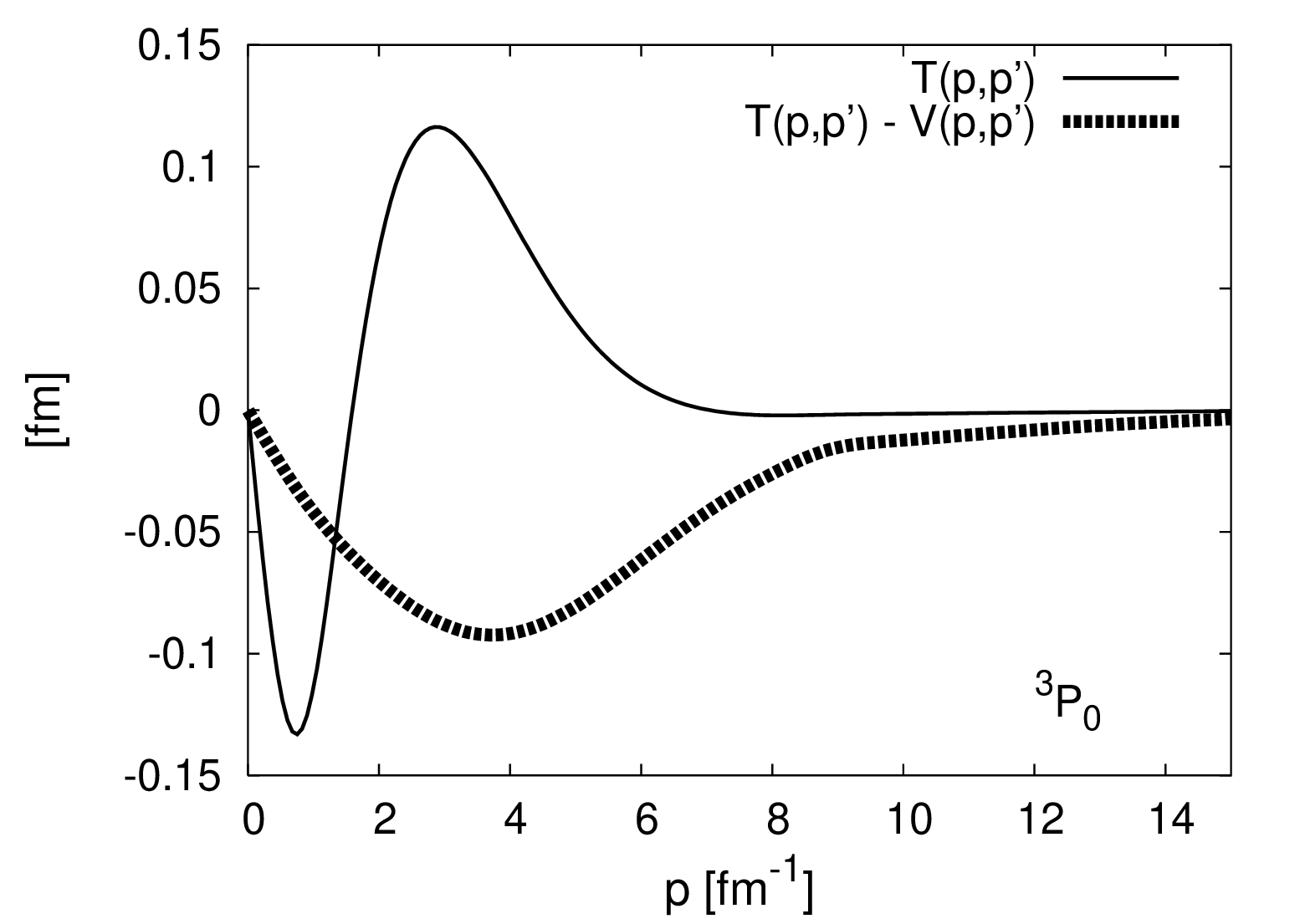}} % postscript image file name
\epsfxsize = 17pc \epsfysize = 9pc 
\centerline{\epsfbox{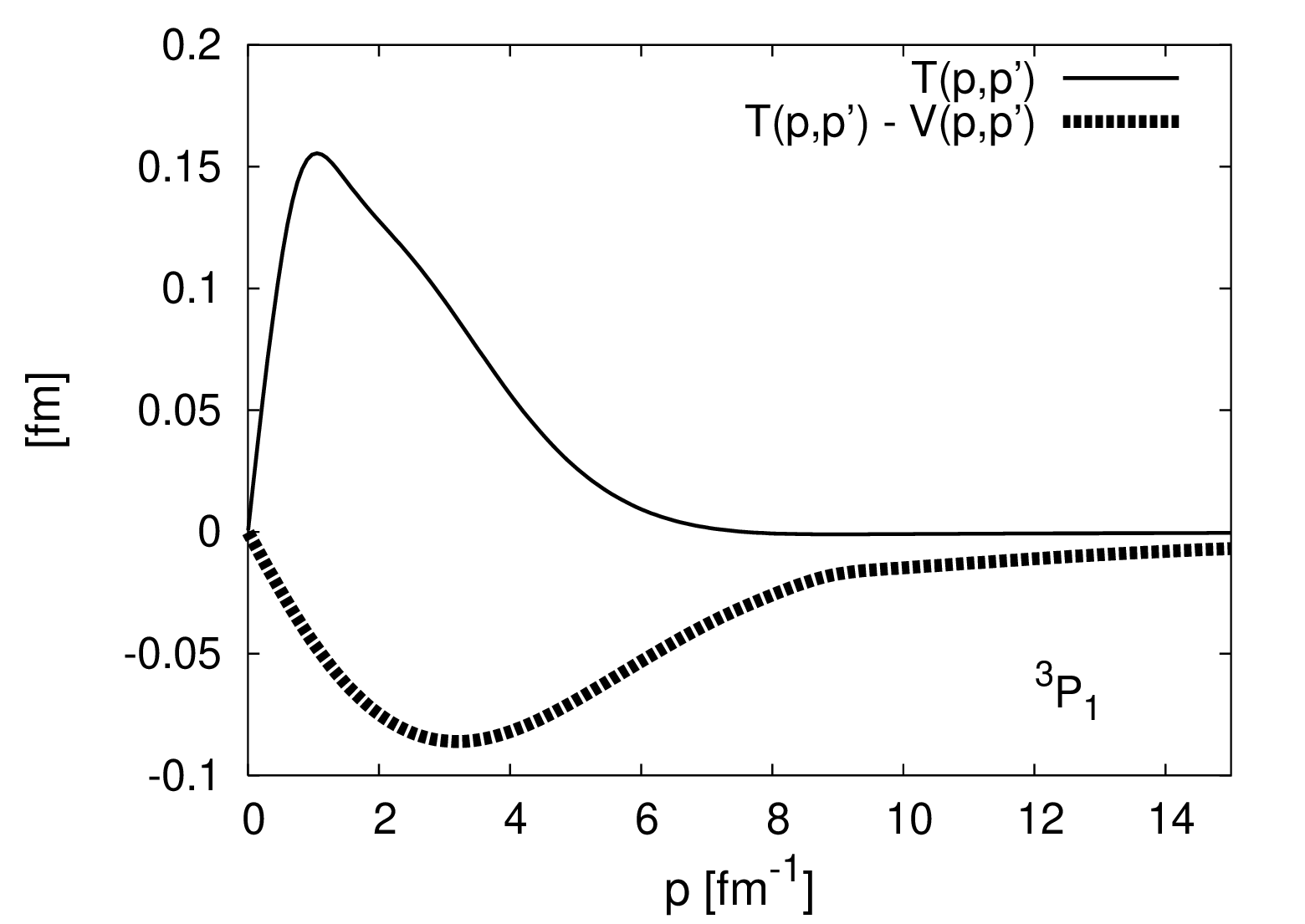}} % postscript image file name
\epsfxsize = 17pc \epsfysize = 9pc 
\centerline{\epsfbox{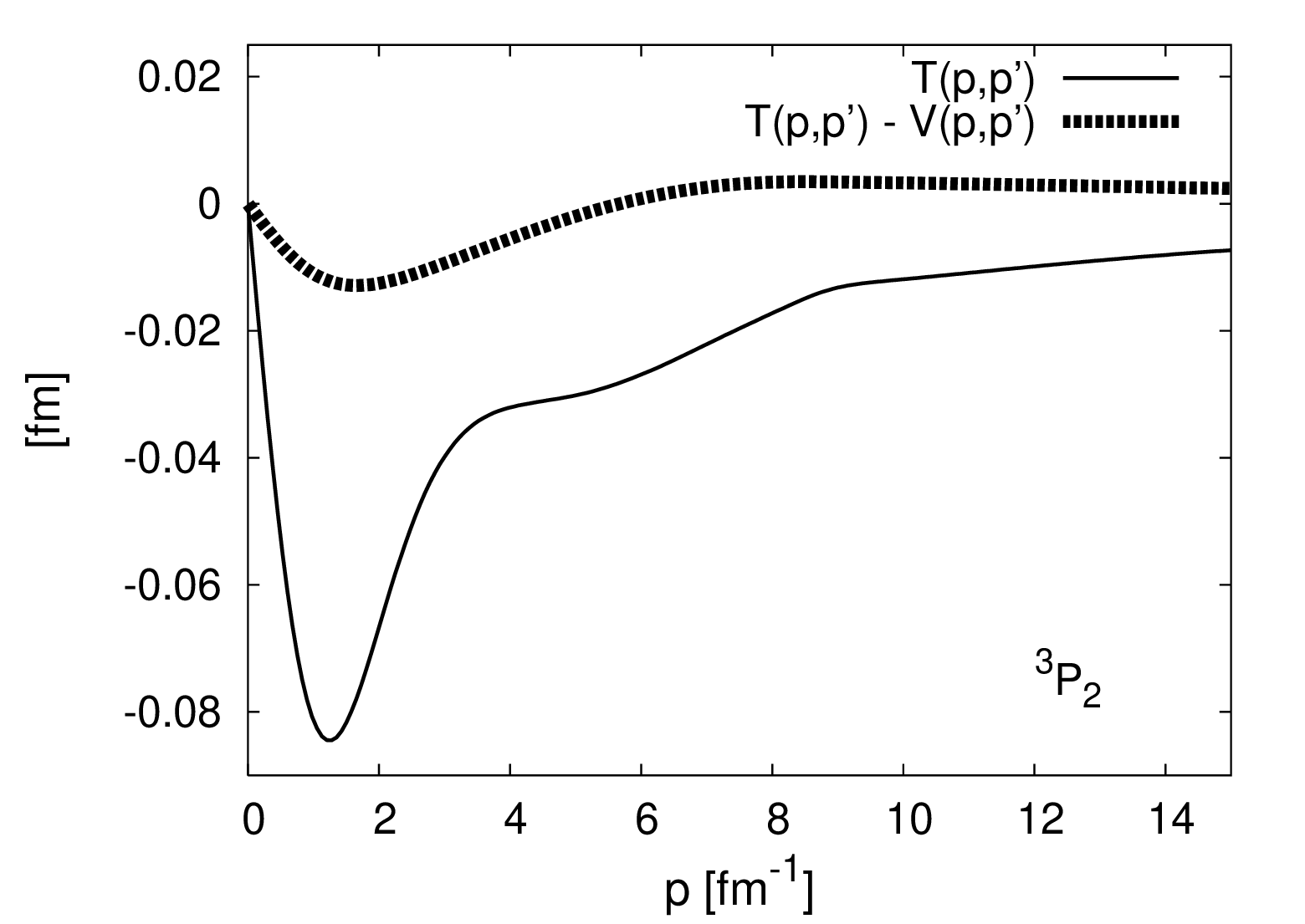}} % postscript image file name
\caption{Comparison between the subtracted (thick-dashed) and 
unsubtracted (thin-solid) $t$-matrix in the triplet $p$-waves for the 
CD-Bonn potential\protect \cite{CD-bonn}. The $t$-matrix has been calculated at
$E=-150$ MeV, and for a fixed momentum $p'$ of 0.089 fm$^{-1}$.}
\label{fig:3}
\end{figure}
%\end{center}
In Fig.~\ref{fig:3} it is shown
that this subtracted $\tilde t$-matrix is not negligible
in the relevant kinematical region, if compared to the 
unsubtracted $t$-matrix. The figure refers to the Bonn CD
potential, however we obtained similar results for other modern 
potentials, and also in other 2$N$ states.

%\begin{center}
\begin{figure}[b]
%\figurebox{20pc}{10pc}{} % to have a box alone
\epsfxsize = 20pc
\epsfysize = 15pc
% will enlarge or reduce the postscript figures based on the xsize
\centerline{\epsfbox{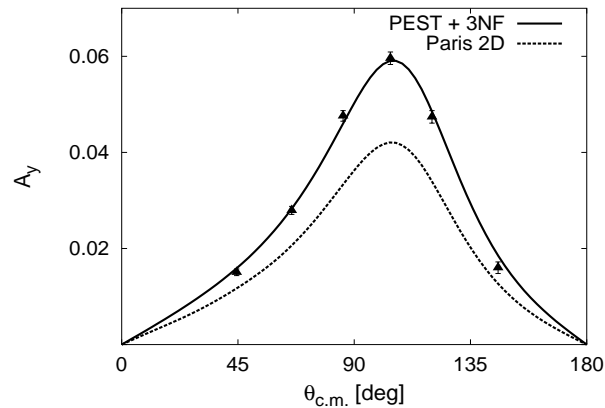}} % postscript image file name
\caption{$A_y$ for $nd$ scattering at 3 MeV (Lab).
Solid line with the tensor $3NF$ herein discussed.
Dashed line without this contribution.
Data from Ref.\protect \cite{Mc}~. }
  \label{fig:4}
\end{figure}
%\end{center}

Fig.~\ref{fig:4} is taken from a recent preprint\cite{CSb} (which
we refer to for all details), and shows the results obtained when the $3NF$
herein discussed is included, with respect to a calculation with
the Paris $2NF$, only. As $2N$ interaction we have considered the 
high-rank parameterization of this interaction known as $PEST$, 
developed by the Graz group\cite{JH}.

\section{Summary}

Pion production reactions have been studied by us theoretically
with a combination of mechanisms involving meson exchanges and $\Delta$
excitations. The results obtained showed that such combination
is appropriate for describing unpolarized and polarized 
observables at low energies.

Also, an approach for the treatment of pion dynamics
has been developed, by taking into account the complete four-body
dynamics of the pion and three nucleon system.
The resulting equations have been treated with an approximation scheme
which leads, to the lowest order, to effective three-nucleon interactions.
It has been shown that one class of diagrams leads to a 
tensor-like three-nucleon potential. It has been argued how
this new term could explain the puzzle of the nucleon-deuteron
vector analyzing powers.

\section*{Acknowledgments}

We thank for support the Italian MURST-PRIN Project 
``Fisica Teorica del Nucleo e dei Sistemi a Pi\'u Corpi". 
We also acknowledge support from the Natural Science and Engineering 
Research Council of Canada. The authors thank also
the Institutions of INFN (Padova), TRIUMF (Vancouver),
University of Manitoba, and Universit\'a di Padova for
hospitality during reciprocal visits.

%\section*{Appendix}
%We can insert an appendix here and place equations so that they
%are given numbers such as Eq.~(\ref{eq:app}).
%\be
%x = y.
%\label{eq:app}
%\ee

\end{document}